\documentclass[aps,pra,twocolumn,amsmath,amssymb,nofootinbib,showpacs,superscriptaddress]{revtex4-2}
\usepackage[english]{babel}
\usepackage{latexsym}
\usepackage{graphics}
\usepackage{graphicx}
\usepackage{epsfig}
\usepackage{color}
\usepackage{bm}
\usepackage{amsmath}
\usepackage{amssymb}
\usepackage{amsthm}
\usepackage{dcolumn}
\usepackage{bm}
\usepackage{float}
\usepackage{hyperref}
\usepackage{color}
\usepackage{epstopdf}
\usepackage{cleveref, braket, comment,bbold}
\usepackage[svgnames]{xcolor}
\hypersetup{hidelinks,colorlinks=true,allcolors=DarkBlue}
\usepackage{ bbold }
\usepackage{tabularx, makecell, booktabs}

\usepackage[normalem]{ulem}

\newcommand{\BLUE}[1]{{\color{black} #1}}

\theoremstyle{remark}

\usepackage{color}
\usepackage{tikz}
\usetikzlibrary{quantikz}
\usepackage{dblfloatfix}

\usepackage{comment}

\begin{document}

\preprint{APS/123-QED}

\title{Generalized Toffoli gate decomposition using ququints: \\ Towards realizing Grover's algorithm with qudits}

\author{A.S. Nikolaeva}
\affiliation{Russian Quantum Center, Skolkovo, Moscow 121205, Russia}
\affiliation{National University of Science and Technology ``MISIS”,  Moscow 119049, Russia}

\author{E.O. Kiktenko}
\affiliation{Russian Quantum Center, Skolkovo, Moscow 121205, Russia}
\affiliation{National University of Science and Technology ``MISIS”,  Moscow 119049, Russia}

\author{A.K. Fedorov}
\affiliation{Russian Quantum Center, Skolkovo, Moscow 121205, Russia}
\affiliation{National University of Science and Technology ``MISIS”,  Moscow 119049, Russia}

\date{\today}
\begin{abstract}
Qubits, which are quantum counterparts of classical bits, are used as basic information units for quantum information processing, whereas underlying physical information carriers, 
e.g. (artificial) atoms or ions, admit encoding of more complex multilevel states -- qudits. 
Recently, significant attention is paid to the idea of using qudit encoding as a way for further scaling quantum processors. 
In this work, we present an efficient decomposition of the generalized Toffoli gate on the five-level quantum systems, so-called ququints, that uses ququints' space as the space of two qubits with a joint ancillary state.
The basic two-qubit operation that we use is a version of controlled-phase gate.
The proposed $N$-qubit Toffoli gate decomposition has $O(N)$ asymptotic depth and does not use ancillary qubits.
We then apply our results for Grover's algorithm, where we indicate on the sizable advantage of the using qudit-based approach with the proposed decomposition in comparison to the standard qubit case. 
We expect that our results are applicable for quantum processors based on various physical platforms, such as trapped ions, neutral atoms, protonic systems, superconducting circuits, and others.
\end{abstract}
\maketitle

\newpage

\section{Introduction}

The concept of quantum computing relies on the idea of manipulating complex (entangled, many-body) 
quantum states in order to solve computational problems that are beyond the capabilities of computing devices based on classical principles~\cite{Manin1980,Feynman1982,Feynman1986}. 
The key problem is, however, to find or engineer a suitable physical platform that allows manipulating and high-efficient control when the system is scaled. 
One of the basic concepts, which is at the heart of the digital quantum computing model~\cite{Deutsch1985}, is to present physical systems as qubits --- two level quantum systems. 
A complexity of defining a general entangled multi-qubit state is exponential in number of qubits;
indeed, a system of $n$ entangled qubits may require up to $2^n$ complex numbers to describe its state~\cite{Brassard1998}
(this is contrast to the classical domain, where a single string of $n$ zeros and ones is sufficient to describe the state of $n$ bits).
This `quantum complexity'~\cite{Preskill2012} can be considered as the origin of quantum computational advantage in solving various problems, such as simulating quantum systems~\cite{Lloyd1996} and prime factorization~\cite{Shor1994}.

Recent experimental progress has been demonstrated with physical platforms of various nature including 
superconducting circuits~\cite{Gambetta2018,Martinis2019,Pan2021-4}, semiconductor quantum dots~\cite{Vandersypen2022,Morello2022,Tarucha2022}, quantum light~\cite{Pan2020,Lavoie2022}, 
neutral atoms~\cite{Lukin2021,Browaeys2021,Browaeys2020-2,Saffman2022}, 
and trapped ions~\cite{Monroe2017,Blatt2012,Blatt2018} (for a review, see Ref.~\cite{Fedorov2022}).
Such setups have been used for testing quantum computational advantage~\cite{Martinis2019,Pan2020,Pan2021-4,Lavoie2022}, 
quantum simulation~\cite{Lukin2021,Browaeys2021,Browaeys2020-2,Monroe2017,Blatt2012}, 
and prototyping various quantum algorithms (see e.g., Refs.~\cite{Gambetta2018,Saffman2022,Blatt2018,Lukin2022}).  
However, computational capabilities of existing prototypes of quantum computing devices are substantially limited. 
The reason behind is the fact that scaling quantum systems with respect to the number of qubits without degrading the quality of control over them remains challenging. 
A clear indication of this fact is that the fidelities of quantum gates in the case of isolated few-qubit systems are much greater than in the case of intermediate-scale systems.
Although, there are no known fundamental obstacles preventing further scaling quantum devices, this task seems to be non-trivial. 
Various approaches, such as new qubit architectures (for example, see Refs.~\cite{Manucharyan2019,Bao2022,Moskalenko2022} for a new type of the currently used superconducting transom qubits~\cite{Gambetta2018,Martinis2019,Pan2021-4}) 
and computational models~\cite{Lloyd2018-2}, have been investigated. 

One may also note that underlying physical platforms for quantum computing, for example, trapped ions and atoms, allow one to encode multiple computational states using a single physical information carrier. 
In other words, such systems can be used for realizing qudit-based quantum processors ($d$-dimensional quantum systems, $d >2$; so the corresponding scaling of the computational space is $d^N$).
The idea of using multilevel (or multistate) quantum systems is known for decades~\cite{Lloyd1993}.
Numerous works on quantum computing with qudits during last decades have confirmed the promise of this approach~\cite{Farhi1998,Kessel1999,Kessel2000,Kessel2002,Muthukrishnan2000,Nielsen2002,Berry2002,Klimov2003,Bagan2003,Vlasov2003,Zobov2008,Zobov2009,Zobov2022,Clark2004,Leary2006,Ralph2007,White2008,Ionicioiu2009,Ivanov2012,Li2013,Kiktenko2015,Kiktenko2015-2, Song2016,Frydryszak2017,Bocharov2017,Gokhale2019,Pan2019,Low2020,Jin2021,Martinis2009,White2009,Wallraff2012,Mischuck2012,Gustavsson2015,Martinis2014,Ustinov2015,Morandotti2017,Balestro2017,Low2020,Sawant2020,Senko2020,Pavlidis2021,Rambow2021,OBrien2022,Nikolaeva2022}.
\BLUE{Besides quantum computing, qudit-based systems offer certain perspectives in quantum teleportation~\cite{Pan2019} and quantum communications~\cite{Gisin2002,Boyd2015}, 
as well as opens up opportunities for uncovering fundamental concepts of quantum mechanics~\cite{Li2013,Frydryszak2017,Zyczkowski2022}.}
It is interesting to note that the first realization of two-qubit gates has used two qubits stored in the degrees of freedom of a single trapped ion, i.e. with the use of a qudit~\cite{Wineland1995}.
One may specifically note demonstrations of multi-qubit processors based on trapped ions~\cite{Ringbauer2021,Semerikov2022}, 
superconducting circuits~\cite{Goss2022,Hill2021,Roy2022}, and optical systems~\cite{OBrien2022}.

A central idea of qudit-based quantum information processing is finding a trade-off between increasing complexity of controlling the system and potential advantages, for example, in realizing quantum algorithms. 
Recent experimental results have demonstrated that it is fairly straightforward to control qudit systems with $d$ up to $7$~\cite{Ringbauer2021,Semerikov2022} with high enough fidelity
using a single laser and acousto-optic modulator.
There are two basic approaches of how additional levels of quantum systems can be used. 
The first idea is to use qudit for substituting ancillas~\cite{Ralph2007,White2009,Ionicioiu2009,Wallraff2012,Gokhale2019,Nikolaeva2022,Kim2023,Nikolaeva2023},
which allows decreasing the number of physical qubits that are required for executing quantum circuits.
Following this method, $N$-qubit Toffoli gate was realized with $2N-3$ qubit-qudit gates on the photonic quantum circuit~\cite{White2009}. 
Although the reduction of the number of qubit-qudit interactions is significant, the experiment configuration, which required $N$-dimensional qudit connected with $N-1$ qubits, is difficult to scale with an increasing number of qubits. 

Another possible approach is to consider qudit’s space as a space of multiple qubits~\cite{Kessel1999,Kessel2000,Kessel2002,Kiktenko2015,Kiktenko2015-2}. 
In this consideration, a reduction in the number of operations can be also achieved, but it depends on the mapping of qubits’ space onto qudits’ space.  
As it has been noted, these approaches can be efficiently combined (this problem has been in general discussed in Ref.~\cite{Nikolaeva2021}).
However, practically relevant cases showing advantages of the combinations of these approaches and their applications in realizing quantum algorithms require additional studies. 

In this work, we study a model of a ququint-based ($d=5$) quantum processor that involves the decomposition of multi-qubit systems in ququint subsystems, as well as the use of ququints’ higher levels as ancillas. 
For this model, we describe possible mapping of qubits’ space onto ququints’ space. 
Then we show how one-qubit gates and the generalized $N$-qubit Toffoli gate can be realized in the proposed setup. 
Finally, we consider how the proposed model is applied to quantum algorithms with a large number of multi-qubit gates. 
The proposed ququint-based quantum processor model allows us to implement $N$-qubit gates with the circuits that have $O(N)$ asymptotic depth.
We apply our results for Grover's algorithm, where we indicate on the sizable advantage of the using qudit-based approach with the proposed decomposition. 

Our work is organized as follows.
In Sec.~\ref{sec:Ququint}, we consider quantum computing with qudits.
Specifically, we analyze a model of a ququint-based ($d=5$) quantum processor that involves the decomposition of multi-qubit systems into ququint subsystems, as well as the use of ququints’ higher levels as ancillas. 
In Sec.~\ref{sec:Toffoli}, we demonstrate how one-qubit gates and the generalized $N$-qubit Toffoli gate can be realized in the proposed setup. 
In Sec.~\ref{sec:Grover}, we present the qudit-based realization of Grover's algorithm.
Finally, we conclude in Sec.~\ref{sec:Conclusion}.

\section{Ququint processor}\label{sec:Ququint}

There are two basic approaches for using additional resources of qudits for realizing more efficient quantum computing. 
We note that the efficiency here can be understood in two aspects. 
The first is that we would like to minimize the number of physical qubits that are used to run quantum circuits. 
Additional complications in realizing quantum circuits comes from the fact that additional ancilla qubits are required, when one would like to implement $N$-qubit gates~\cite{Barenco1995}. 
For example, efficient implementation of $N$-qubit Toffoli gates is essential for Grover search~\cite{Grover1996,Grover1997}.
One can then use additional levels of qudits for substituting ancillas~\cite{Ralph2007,White2009,Ionicioiu2009,Wallraff2012,Gokhale2019,Nikolaeva2022,Kim2023,Nikolaeva2023},
which allows decreasing the number of physical qubits that are required for executing quantum circuits.

Another idea is to consider qudit’s space as a space of multiple qubits~\cite{Kessel1999,Kessel2000,Kessel2002,Kiktenko2015,Kiktenko2015-2}. 
A reduction in the number of operations can be also achieved, but it depends on the mapping of qubits’ space onto qudits’ space.  
As it has been noted, these approaches can be efficiently combined (this problem has been in general discussed in Ref.~\cite{Nikolaeva2021}).
However, practically relevant cases showing advantages of the combinations of these approaches and their applications in realizing quantum algorithms require additional studies. 
Below we consider an important particular case for combining these approaches.

\subsection{Ququint as two qubits and ancillary state}

Five-dimensional state space of a ququint $Q$ can be considered as a joint space of two qubits, $a$ and $b$, accompanied with an ancillary state.
The corresponding qubit-to-qudit mapping can be represented as follows:
\begin{equation} \label{eq:mapping}
\begin{aligned}
	\ket{0}_Q&\to \ket{0}_a\otimes\ket{0}_b,\\
	\ket{1}_Q&\to \ket{0}_a\otimes\ket{1}_b,\\
	\ket{2}_Q&\to \ket{1}_a\otimes\ket{0}_b,\\
	\ket{3}_Q&\to \ket{1}_a\otimes\ket{1}_b,\\
	\ket{4}_Q&\to \ket{\rm anc},\\
\end{aligned}
\end{equation}
where $\ket{n}_Q$ with $n=0,\ldots,4$ denote basis states of ququint $Q$, $\ket{m}_{a(b)}$ with $m=0,1$ denote computational basis states of qubit $a (b)$, which is embedded in $Q$, and $\ket{\rm anc}$ denotes the ancillary state. 
Below we assume that the state $\ket{\rm anc}$ serves only as a `pure' ancilla for implementing multi-qubit gates: According to the designed decomposition, this level is populated only during the realization of a multi-qubit gate
(it is initialized in the state $\ket{0}$ and it is also in this state at the final step of implementing quantum circuits).
The introduced representation of ququint's space allows one to reduce the required number of physical systems and two-qudit gates in multi-qubit gate decomposition, as we demonstrate in the next section.

\begin{figure}[h]
	\begin{center}
\includegraphics[width = 0.8\linewidth]{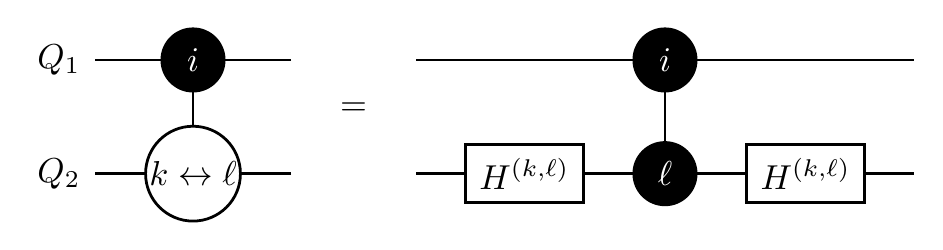}
		\end{center}
    \caption{Realization of a generalized controlled inversion ${\sf CX}^{i\rightarrow k,\ell}_{Q_1Q_2}$ gate via generalized controlled-phase ${\sf CZ}^{i\leftrightarrow j}_{Q_1Q_2}$ and $H^{(k,\ell)}_{Q_2}$ gate on two ququints.
    \BLUE{
    In the left hand side of the identity, the black-painted circle with white $i$ denotes a control qudit and the control state $\ket{i}$ for ${\sf CX}^{i\rightarrow k,\ell}_{Q_1Q_2}$ gate. The corresponding target qudit is denoted by the white circle with an arrow between $k$ and $\ell$.
    ${\sf CX}^{i\rightarrow k,\ell}_{Q_1Q_2}$ gate exchanges populations between levels $\ket{k}$  and $\ket{\ell}$ state of the target qudit, given that control qudit is in the state $\ket{i}$.
    In the right hand side of the identity, the gate with two connected black-painted circles corresponds to  controlled-phase ${\sf CZ}^{i\leftrightarrow \ell}_{Q_1Q_2}$ operation, which applies a phase factor $-1$ to the state of two ququints $\ket{i\ell}_{Q_1Q_2}$, and left other states unchanged. 
    Single qudits $H^{(k,\ell)}$ gates denote two-dimensional Hadamard transformations realized at levels $\ket{k}$ and $\ket{\ell}$ of $Q_2$.
    }}
    \label{fig:CX_ikl}
\end{figure}

In analogy with the idea of qubit-based digital quantum computations (see above), 
we assume that we can perform any desirable single-qubit unitary operation $U\in \mathbf{U}(2)$ on an arbitrary pair of levels $i$ and $j$ in ququint $Q$. 
The resulting unitary operation, denoted $U^{(i,j)}_Q$, takes the following form:
\begin{multline}
	U^{(i,j)}_Q = \bra{0} U \ket{0} \ket{i}_Q\bra{i}+
	\bra{0} U \ket{1} \ket{i}_Q\bra{j} \\
	+\bra{1} U \ket{0} \ket{j}_Q\bra{i}+
	\bra{1} U \ket{1} \ket{j}_Q\bra{j}+
	\mathbb{1}_{i,j}^\perp,
\end{multline}
where $\mathbb{1}_{i,j}^\perp$ stands for a projector on a three-dimensional orthogonal complement of the subspace spanned by $\ket{i}_Q,\ket{j}_Q$.
According to mapping~\eqref{eq:mapping}, applying single-qubit gate
\begin{equation}
	U = 
	\begin{pmatrix}
		\alpha & \beta\\
		\gamma & \delta
	\end{pmatrix}
\end{equation}
to qubits $a$ and $b$, respectively, leads to the following single-qudit realizations:
\begin{equation}
\begin{aligned}
	U_a &\equiv U^{(0,2)}_{Q} U^{(1,3)}_{Q}=
	\begin{pmatrix}
        \alpha & & \beta & &\\
        &\alpha & & \beta  &\\
        \gamma & &\delta & & \\
        &\gamma & & \delta  &\\
        & &  & & 1
	\end{pmatrix}\\
	U_b &\equiv U^{(0,1)}_{Q} U^{(2,3)}_{Q}=
	\begin{pmatrix}
        \alpha & \beta & & &\\
        \gamma &\delta & &   &\\
        & &\alpha & \beta  & \\
	& &\gamma & \delta  & \\
	& &  & & 1
	\end{pmatrix}
\end{aligned}
\end{equation}
(here and after all unspecified elements in matrices are zeros).

One of the main features of the considering the ququint’s space as a space of two qubits with ancillary level is an ability to implement two-qubit gates between qubits $a$ and $b$ using single-ququint gates only. 
A controlled-phase gate between $a$ and $b$ can be realized with a single-qudit operation 
\begin{equation}
	{\sf CZ}_{ab} \equiv {\sf Z}_Q^{(0,3)}=
	\begin{pmatrix}
	1 & & & &\\
		& 1 & & &\\
		& & 1 &  & \\
		& & & -1 & & \\
		& &  & & 1
	\end{pmatrix},
\end{equation}
where ${\sf Z}=\ket{0}\bra{0}-\ket{1}\bra{1}$ is a standard Pauli matrix.
We note that other realizations, e.g., ${\sf Z}_Q^{(1,3)}$ are possible.

We note that any restricted, yet connected, coupling map between levels inside a qudit is enough to implement unitary operation on arbitrary pair of levels~\cite{Mato2022, Low2020}.
For example, in order to couple levels $\ket{0}_Q$ and $\ket{2}_Q$, 
one can use transitions $\ket{0}_Q\leftrightarrow \ket{1}_Q$ and $\ket{1}_Q\leftrightarrow \ket{2}_Q$, even in the case, where transition $\ket{0}_Q\leftrightarrow \ket{2}_Q$ is forbidden due to selection rules.
Moreover, in actual existing experimental setups, transitions within a given coupling graph can be implemented with a single laser and acousto-optic modulator~\cite{Ringbauer2021}.

As a two-ququint gate we consider ${\sf CZ}^{i\leftrightarrow j}_{Q_1Q_2}$ gate, which applies phase factor $-1$ to the state $\ket{ij}_{Q_1Q_2}$ of two ququints $Q_1$ and $Q_2$:
\begin{equation}
	{\sf CZ}^{i\leftrightarrow j}_{Q_1Q_2}=\sum_{m,n}(-1)^{\delta_{i,m}\delta_{j,n}}\ket{m}_{Q_1}\bra{m}\otimes\ket{n}_{Q_2}\bra{n}.
\end{equation}
We note that this two-qudit gate can be realized via Rydberg blockade neutral atom-based \cite{Gonzalez2022} qudits, and via common quantized motion mode in ion-based platform \cite{Cirac1995}.
On the basis of ${\sf CZ}^{i\leftrightarrow j}_{Q_1Q_2}$ and $U^{(i,j)}_{Q_{2(1)}}$ gates one can construct more complicated gates, such as ${\sf CX}^{i\rightarrow k,\ell}_{Q_1Q_2}$ gate, which is defined as:
\begin{equation}
    {\sf CX}^{i\rightarrow k,\ell}_{Q_1Q_2} = H^{(k,\ell)}_{Q_2} {\sf CZ}^{i\leftrightarrow \ell}_{Q_1 Q_2} H^{(k,\ell)}_{Q_2},
\end{equation}
where ${\sf H}=2^{-1/2}\sum_{p,r=0,1}(-1)^{pr}\ket{p}\bra{r}$ is a standard Hadamard gate (see also Fig.~\ref{fig:CX_ikl}).
The idea of two-ququint ${\sf CX}^{i\rightarrow k,\ell}_{Q_1Q_2}$ gate is similar to the the idea of qubit $\sf CX$ gate: 
It swaps states $\ket{k}_{Q_2}$ and $\ket{\ell}_{Q_2}$ when $Q_1$ is in the state $\ket{i}_{Q_1}$. 

To conclude this section, we note that each ququint $Q$ can be also used for embedding a single qubit $a$ accompanied with three ancillary levels.
In this case, the qubit-to-qudit mapping takes the form
\begin{equation} \label{eq:mapping-1-qubit}
\begin{aligned}
	\ket{0}_Q&\to \ket{0}_a,\\
	\ket{1}_Q&\to \ket{1}_a,\\
	\ket{2}_Q&\to \ket{\rm anc},\\
	\ket{3}_Q&\to \ket{\rm anc'},\\
	\ket{4}_Q&\to \ket{\rm anc''},\\
\end{aligned}
\end{equation}
where $\ket{\rm anc'}$ and $\ket{\rm anc''}$ denote new auxiliary levels.

We assume that each qudit can be measured in the computational basis.
From the viewpoint of embedded qubit(s), this measurement corresponds to the computational basis measurement over one or two qubits. 
The correspondence is given by mapping~\eqref{eq:mapping} or \eqref{eq:mapping-1-qubit}.
Below we consider both mappings~\eqref{eq:mapping} and~\eqref{eq:mapping-1-qubit} within decomposition of the generalized $N$-qubit Toffoli gate.

\section{Toffoli gate implementation}\label{sec:Toffoli}

The generalized $N$-qubit Toffoli gate ${\sf C}^{N-1}{\sf X}^{(t)}$, acting on qubits $q_1, \ldots, q_N$, 
flips a particular target qubit state of $q_t$ if and only if all the other $N-1$ control qubits are in the state $\ket{1}$~\cite{Barenco1995}.  
This gate can be realized with `symmetric' multi-controlled phase gate
\begin{multline} \label{eq:global_CZ_gate}
    {\sf C}^{N-1}{\sf Z}\ket{b_1\ldots b_N}_{q_1\ldots q_N} \\= (-1)^{b_1\ldots b_N} \ket{b_1\ldots b_N}_{q_1\ldots q_N},
\end{multline}
where $b_i=0,1$ denote qubit basis states.
One can obtain ${\sf C}^{N-1}{\sf X}^{(t)}$~ from ${\sf C}^{N-1}{\sf Z}$ by surrounding the target qubit $t$ with single-qubit Hadamard gates.
In what follows we consider a ququint-based implementation ${\sf C}^{N-1}{\sf Z}$.

To clarify our consideration, we discuss the most simple cases of embedding qubits that are effected by ${\sf C}^{N-1}{\sf Z}$ gate in qudits.
For even $N$, we consider $N$ qubits embedded in $N/2$ ququints according to mapping~\eqref{eq:mapping}.
For odd $N$, we consider $N-1$ qubits embedded in $(N-1)/2$ ququints, and the remaining qubit embedded in an additional ququint.
Two situations are possible: (i) the additional ququint is used for storing the single $N$th qubit only [i.e., mapping~\eqref{eq:mapping-1-qubit} is used], and (ii) there exists an additional neighboring qubit, embedded in this ququint, that is involved in the whole qubit circuit, but not involved in the decomposed ${\sf C}^{N-1}{\sf Z}$ gate [mapping~\eqref{eq:mapping} is used].
These three cases require separate treatments (see Tab. \ref{tab:gates}).

Let us start with a decomposition of ${\sf C}^2{\sf Z}$ gate ($N=3$).
For this purpose, we embed qubits $q_1$ and $q_2$ in the single ququint $Q_1$ and consider two variants of embedding $q_3$ in $Q_2$.
If $q_3$ is embedded in $Q_2$ according to mapping~\eqref{eq:mapping-1-qubit}, the implementation of ${\sf C}^2{\sf Z}$ reduces to realization of ${\sf CZ}^{(3\leftrightarrow 1)}_{Q_1 Q_2}$ gate, 
since $\ket{31}_{Q_1Q_2}$ is mapped to $\ket{111}_{q_1q_2q_3}$.
In the case of mapping~\eqref{eq:mapping} for $Q_2$ [we assume that $q_2$ corresponds to $a$ in~\eqref{eq:mapping}], we apply two gates: ${\sf CZ}^{3\leftrightarrow 2}_{Q_1 Q_2}$ and ${\sf CZ}^{3\leftrightarrow 3}_{Q_1 Q_2}$.
This overhead in the number of gates is due to necessity to preserve the state of the second qubit, embedded in $Q_2$.
We note, that this doubling of the number of gates meets us in every decomposition of ${\sf C}^{N-1}{\sf Z}$ with odd $N$ and appearance of neighbouring qubit in the last $[(N+1)/2]$th ququint.

\begin{table}[]
\begin{center}
\begin{tabularx}{\linewidth}{ | >{\arraybackslash\centering}m{0.6cm} | >{\centering} X| >{\centering}X|}
   \hline
 \multicolumn{1}{|p{0.6cm}|}{\vspace*{0.55cm} ${\sf C}^2{\sf Z}$}
		& {mapping~(\ref{eq:mapping-1-qubit}) for $Q_{2}$}
			\includegraphics[width=0.46\linewidth]{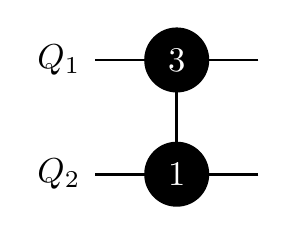}
		&  {mapping~(\ref{eq:mapping}) for $Q_{2}$}
			\includegraphics[width=0.65\linewidth]{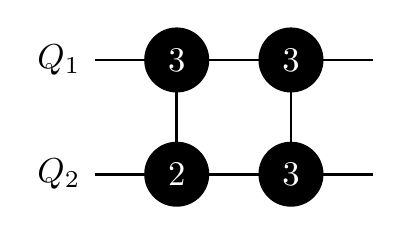}
        \tabularnewline \hline
		\multicolumn{1}{|m{0.6cm}|}{${\sf C}^3{\sf Z}$\vspace*{0.45cm}}& \multicolumn{2}{>{\centering}c|}{
			\includegraphics[width=0.2\linewidth]{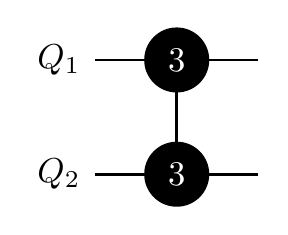}\phantom{\hspace{0.4cm}}}
	\\ \hline
		\multicolumn{1}{|p{0.6cm}|}{\vspace*{0.8cm}${\sf C}^4{\sf Z}$}
	&{mapping~(\ref{eq:mapping-1-qubit}) for $Q_{3}$}
		\includegraphics[width=0.8\linewidth]{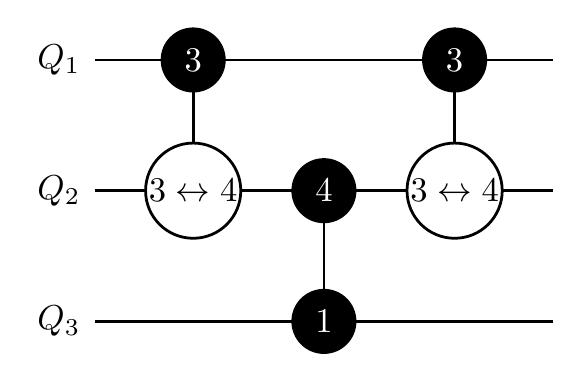}
	& {mapping~(\ref{eq:mapping}) for $Q_{3}$}
		\includegraphics[width=0.95\linewidth]{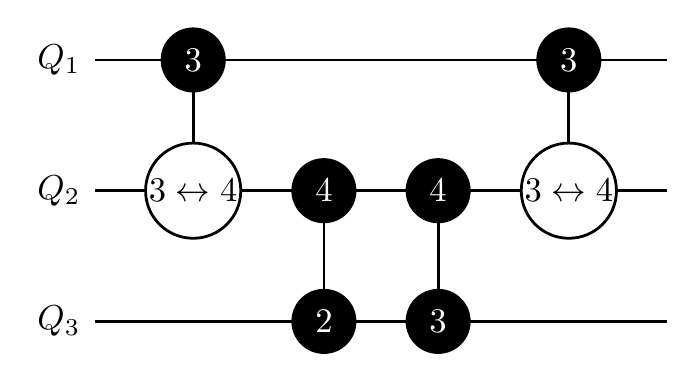}
	\tabularnewline \hline
	\multicolumn{1}{|m{0.6cm}|}{${\sf C}^5{\sf Z}$\vspace*{0.75cm}}& \multicolumn{2}{c|}{
			\includegraphics[width=0.35\linewidth]{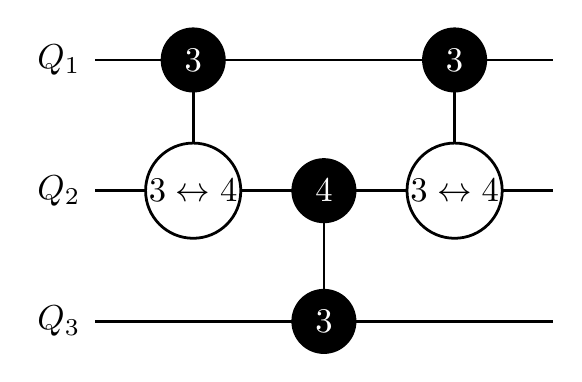}\phantom{\hspace{0.35cm}}
	} 
	\\ \hline
	\end{tabularx}
 \end{center}
	\caption{${\sf C}^{N-1}{\sf Z}$ gate implementation on ququints for $N = 3,\dots, 6$. with  ${\sf CZ}^{i\leftrightarrow j}$ and ${\sf CX}^{i\rightarrow k,\ell}$ gates for two-possible variants of mapping for the `bottom' ququint.
 \BLUE{
  Labeling of gates is the same as in Fig.~\ref{fig:CX_ikl}.
 }
}
 \label{tab:gates}
\end{table}

Following this idea, the implementation of ${\sf C}^{3}{\sf Z}$ on two ququints $Q_1$ and $Q_2$ is realized via ${\sf CZ}^{3\leftrightarrow 3}_{Q_1 Q_2}$ gate.

To implement a five-qubit ${\sf C}^{4}{\sf Z}$ gate we use the following trick.
We put the information about whether the four qubits $q_1,\ldots,q_4$, embedded in $Q_1$ and $Q_2$, are in unit state $\ket{1\ldots1}_{q_1\ldots q_4}$ in the ancillary state of $Q_2$.
It can be realized by applying ${\sf CX}^{3\rightarrow 3,4}_{Q_1Q_2}$ gate.
Then we apply the controlled-phase rotation from the ancillary state of $Q_2$ to the state of $q_5$, embedded in $Q_3$.
Depending on the type of a mapping used for $Q_3$, we apply a single two-ququint gate ${\sf CZ}^{4\leftrightarrow 1}_{Q_2 Q_3}$ or two two-ququint gates ${\sf CZ}^{4\leftrightarrow 2}_{Q_2 Q_3}$ and ${\sf CZ}^{4\leftrightarrow 3}_{Q_2 Q_3}$.
Note that the phase factor is acquired if and only if all five qubit are initially in the unit state.
At the final step, the state $Q_1$ and $Q_2$ is restored to the original state by `uncomputation' with ${\sf CX}^{3\rightarrow 3,4}_{Q_1Q_2}$ gate.
One can see that key idea of this decomposition is that we store information about two qubits in the first four ququint levels, and the highest ququint state $\ket{4}\equiv\ket{\rm anc}$ is used instead of an ancilla to store temporary data.

\begin{figure}[h]
	\centering
	\includegraphics[width = 0.95\linewidth]{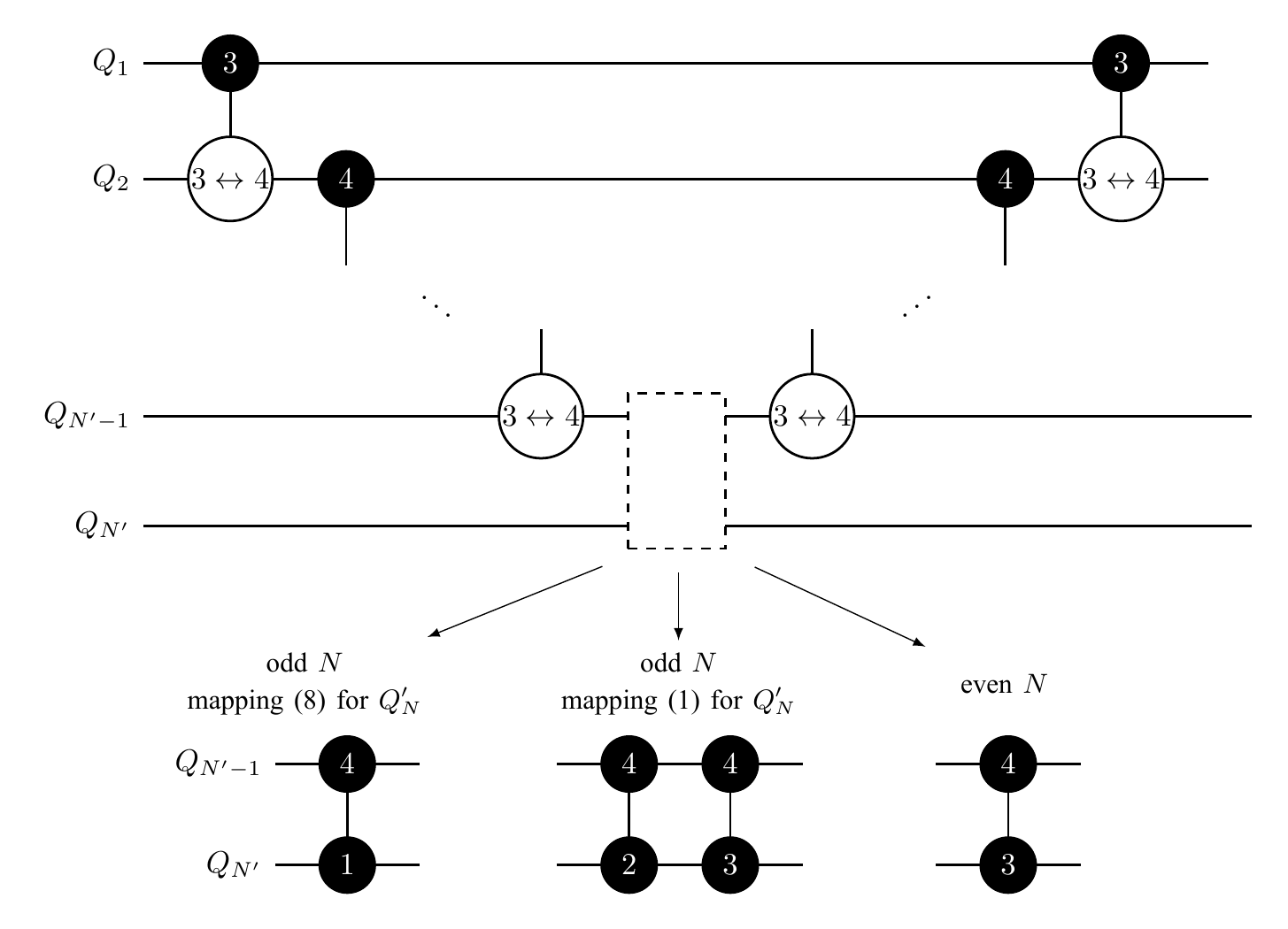}
	\caption{${\sf C}^{N-1}{\sf Z}$ gate decomposition  on ququints with ${\sf CX}^{i\rightarrow k,\ell}$ gates for $N \geq 6$. 
	In the central part of the circuit  we apply a controlled-phase gate
	${\sf CZ}^{4\leftrightarrow 1}_{Q_{N'-1} Q_{N'}}$ if $N$ is odd and mapping~\eqref{eq:mapping-1-qubit} is used, or two gates ${\sf CZ}^{4\leftrightarrow 2}_{Q_{N'-1} Q_{N'}}$ 
	and ${\sf CZ}^{4\leftrightarrow 3}_{Q_{N'-1} Q_{N'}}$ if $N$ is odd and 		mapping~\eqref{eq:mapping} is used, or a controlled-phase gate
	${\sf CZ}^{4\leftrightarrow 3}_{Q_{N'-1} Q_{N'}}$ if $N$ is even.
  \BLUE{
  Labeling of gates is the same as in Fig.~\ref{fig:CX_ikl}.
  }
 }
	\label{fig:gen_cnz}
\end{figure}

The decomposition of ${\sf C}^{4}{\sf Z}$ allows us to obtain a decomposition for ${\sf C}^{N-1}{\sf Z}$ gate with $N \geq 6 $ (see Fig. \ref{fig:gen_cnz}).
The key idea is the same.
We apply a sequence
${\sf CX}^{3\rightarrow 3,4}_{Q_1 Q_2}$, ${\sf CX}^{4\rightarrow 3,4}_{Q_2 Q_3}$, \ldots,  ${\sf CX}^{4\rightarrow 3,4}_{Q_{N'-2} Q_{N'-1}}$, where $N'=N/2$ for even $N$ and $N'=(N+1)/2$ for odd $N$.
It brings $(N'-1)$th qudit in the ancillary state if and only if all qubits embedded in $Q_1,\ldots,Q_{N'}$ are in unit sate.
Then we apply a controlled-phase gate
${\sf CZ}^{4\leftrightarrow 3}_{Q_{N'-1} Q_{N'}}$ if $N$ is even, or a controlled-phase gate
${\sf CZ}^{4\leftrightarrow 1}_{Q_{N'-1} Q_{N'}}$ if $N$ is odd and mapping~\eqref{eq:mapping-1-qubit} is used, or two gates ${\sf CZ}^{4\leftrightarrow 2}_{Q_{N'-1} Q_{N'}}$ and ${\sf CZ}^{4\leftrightarrow 3}_{Q_{N'-1} Q_{N'}}$ if $N$ is odd and mapping~\eqref{eq:mapping} is used.
Finally, the ladder of ${\sf CX}$ gates is implemented in reverse order.
In the result we obtain a circuit consisted of $N-3$ or $N-2$ two-ququint gates that has $O(N)$ asymptotic depth.

\begin{figure}[]
\includegraphics[width=0.95\linewidth]{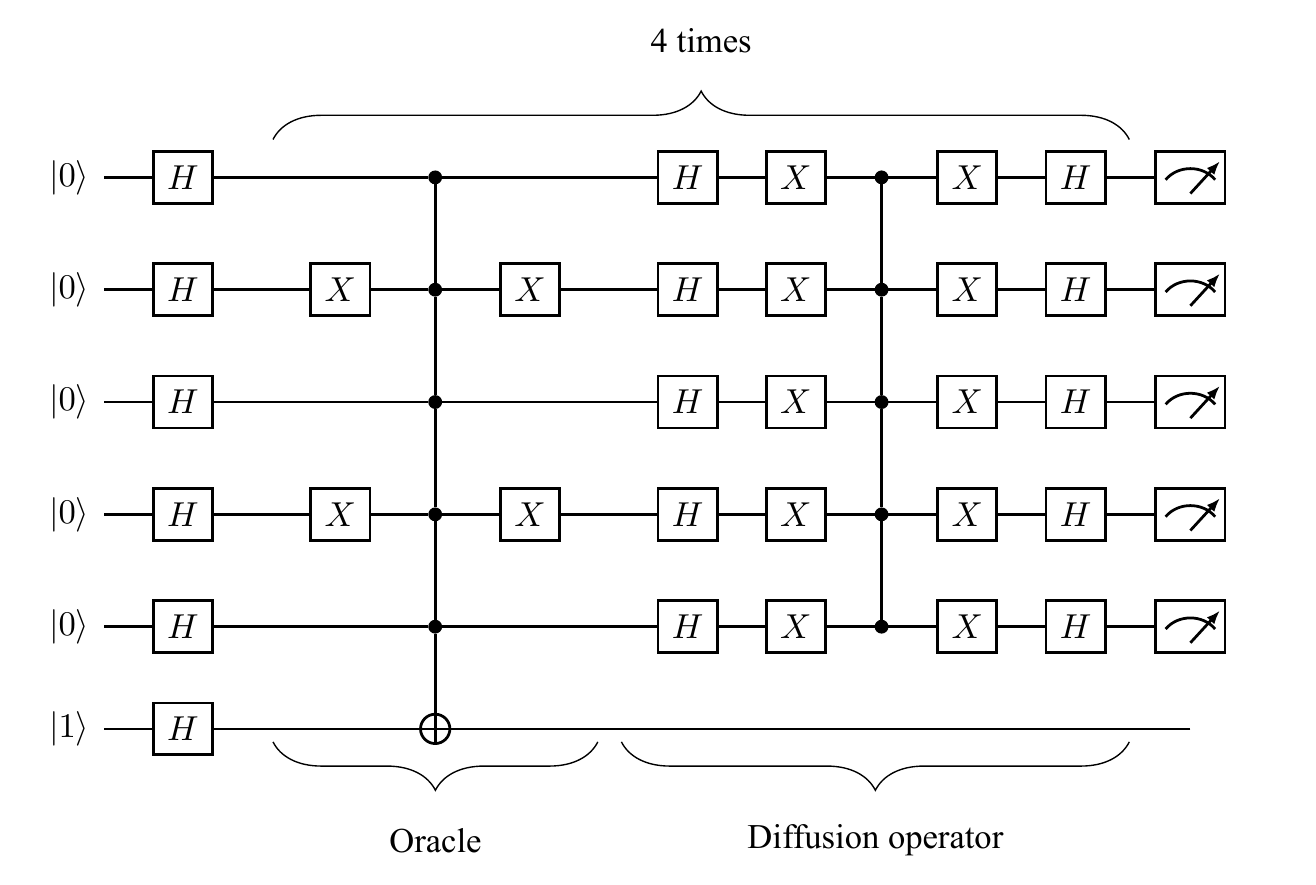}
\caption{Grover's algorithm for search item $\omega = 10101$ over $2^5=32$ items. 
Each of four iterations has two multiply-controlled gates: one in the oracle and one in the diffusion operator. Both these multiply-controlled gates can be efficiently decomposed into two-qudit gates with ququints.}
    \label{fig:grover}
\end{figure}

\begin{figure}[h!]
    \centering
    \includegraphics[width=0.9\linewidth]{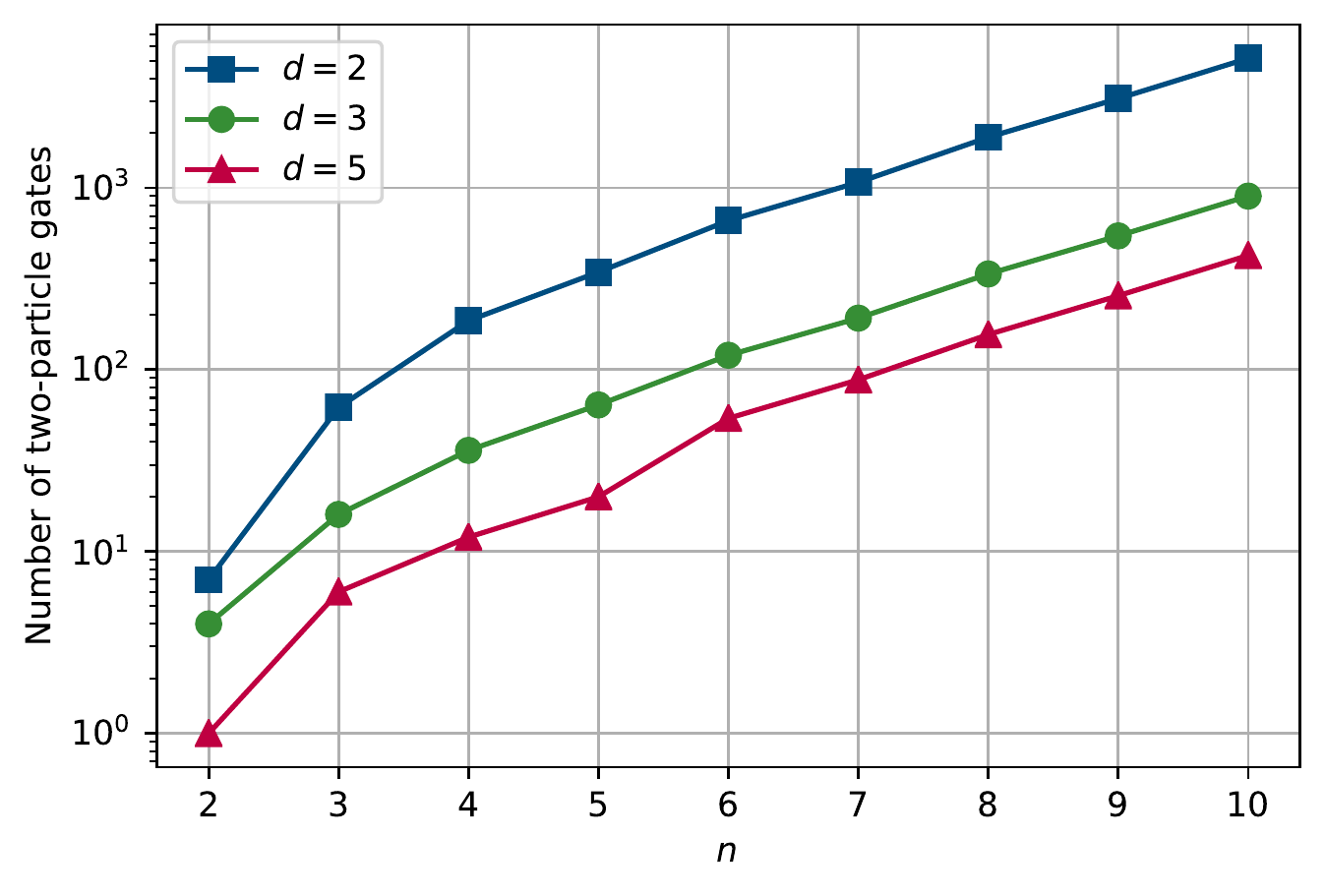}
    \caption{Two-qudit gate counts for implementations of $n$-qubit Grover's algorithm ($n$ is from 2 to 10) with qubit-based decomposition method~\cite{Nielsen2002}, which requires $n-2$ ancillary qubits for $n$-qubit gate decomposition, and has linear scaling, 
    qutrit-based decomposition method~\cite{Nikolaeva2022}, and the proposed ququint-based decomposition method.
    \BLUE{Plotted data takes into account an increase in the number of Grover's steps in quantum circuits with an increase in the number of involved qubits.}
    }
    \label{fig:gates}
\end{figure}

\section[2]{Application to Grover's algorithm}\label{sec:Grover}

The method proposed in the present work to construct the generalized Toffoli gate can be applied to any quantum algorithm that contains multi-qubit gates. 
A clear example is Grover's algorithm~\cite{Grover1996,Grover1997} for searching a `hidden' bitstring $\omega\in\{0,1\}^n$, s.t. $f(\omega)=1$, 
where a `black box' function $f: \{0,1\}^n\rightarrow \{0,1\}$ is known to take a unit value only on one element.
Here $n$ is some integer value, which defines a domain for $f$ and determines complexity of the problem.
Grover's algorithm typically requires $O(2^{n/2})$ queries to an oracle $U_f: \ket{x}\ket{t}\mapsto \ket{x}\ket{f(x)\oplus t}$, where $x\in\{0,1\}^n$, $t\in\{0,1\}$, and $\oplus$ stands for exclusive OR (XOR) operation.
Each query is followed by a $n$-qubit diffusion operator 
\begin{equation}
    D=\mathbb{1}-2\ket{\rm sym}\bra{\rm sym},
\end{equation}
where $\mathbb{1}$ is $2^n$-dimensional identity matrix, and $\ket{\rm sym}:=2^{-n/2}\sum_{x\in\{0,1\}^n}\ket{x}$.
Notably, diffusion operator can be reduced to ${\sf C}^{n-1}{\sf Z}$ gate surrounded by single-qubit ${\sf H}$ and ${\sf X}=\ket{0}\bra{1}+\ket{1}\bra{0}$ gates.
The explicit circuit diagram for finding item $\omega = 10101$ $(n=5)$ items is shown in Fig.~\ref{fig:grover}.

\BLUE{To examine an efficiency of the proposed decomposition, we compute a two-particle gates count for Grover's algorithm implementations with several approaches for the decomposition of multi-qubit gates, where by a `two-particle gate' we mean a gate between two physical systems, which are used as qudits. This term is convenient to use as we further compare total amount of operations between physical systems with different number of levels in the implementation of Grover's algorithm.}
\BLUE{For the comparison we chose three approaches to the decomposition of multi-qubit gates: }
qubit-based decomposition with additional ancillary qubits~\cite{Nielsen2002}, qutrit based decomposition, where higher levels of qutrits act as ancillas~\cite{Kiktenko2020,Nikolaeva2022}, and ququint-based decomposition that is proposed above.

\BLUE{
For all described decomposition methods, two-qudit gate counts, resulting from the implementation of Grover's algorithm on 2 to 10 qubits, are plotted in Fig.~\ref{fig:gates}.
We note, that the plotted data takes into account an increase in the number of iterations (Grover's step) increase in the number of involved qubits. 

The first considered method \cite{Nielsen2002} relies on using only qubits for the decomposition of multi-qubit gate (square symbols line in Fig.~\ref{fig:gates}). 
For this reason, to achieve linear scaling of the required number of two-qubit gates in the decomposition, it is necessary to use additional qubits as ancillas.
Namely, to decompose $n$-qubit controlled gate, one needs to use $n-2$ additional physical systems.
Using them, the number of required two-qubit gates to implement $n$-qubit gate is equal to $12n-23$. 

Qutrit-based decomposition \cite{Kiktenko2020,Nikolaeva2022} provides significantly lower constant in a linear scaling of the required number of two-particle gates in the $n$-qubit gate decomposition (circle symbols in Fig.~\ref{fig:gates}). 
There is no need to use additional physical qubits within this decomposition, owing to the presence of the third level in qutrit and its use as an ancillary state.
The main idea behind qutrit-based decomposition is to ``check'' the states of each pair of qubit sequentially, leaving the second qutrit in the pair in state $\ket{1}$ if and only if both qutrits are in the state $\ket{1}$. 
Then, if this condition is satisfied, required controlled operation ($\sf CZ$ or $\sf CX$) is applied to the last pair of qutrits. 
It can be seen that the circuit of this decomposition has V-ladder-like architecture and consists of $2n-3$ two-particle gates for $n$-qubit gate.

The third decomposition, which is considered for comparison, is the proposed in this paper ququint-based decomposition (triangle symbols in Fig.~\ref{fig:gates}). 
As it was discussed earlier, the main feature of this method is that the reduction in the number of required two-particle gates is provided by embedding two qubits in a single ququint the together with the use of the highest level in ququint as an ancillary state. 
Such combined approach to the use of ququints' space make the constant in the linear scaling number of two-particle gates even lower than in qutrit-based decomposition. }

As we can see from Figure~\ref{fig:gates}, the use of ququints allows us to reduce the number of two-qudit gates in the implementation of Grover's algorithm 
by a thousand times compared to its only-qubit implementation if the number of items for search is greater than $n=8$. 
However, on a small number of required qubits, the ququint-based method and the method from Ref.~\cite{Kiktenko2020} have almost the same efficiency. 
The reason for this is that the contribution from information compression from two qubits in one physical ququint grows with the number of required qubits in the algorithm. 
For this reason, the ququint-based decomposition of multi-qubit gates is optimal for algorithms with a sufficiently large ($n>4$) number of required qubits.

\section{Conclusion \& outlook}\label{sec:Conclusion}

We have demonstrated that a sizable reduction in the number of gates in the quantum circuit can be achieved by considering ququint's space as two qubits and a joint ancillary state.  
We have presented a new decomposition of the generalized $N$-qubit Toffoli gate that  uses  no additional ancillary qubits  and requires a linear number of two-particle gates. 
The new decomposition can be exploited in algorithms with multi-qubit gates and error correction schemes to increase the total circuit fidelity. 
Its efficiency we demonstrate on the Grover's search algorithm, which is a good illustrative example since it requires multi-qubit gates in both parts an oracle, and the diffusion operator of the algorithm. 
The crucial point is that the resulting number of two-particle gates required for implementing its circuit with ququints appears considerably smaller than the one in a straightforward qubit-based implementation.

In the current contribution we have considered a general theoretical approach, leaving a detailed design for particular physical platforms for future works. 
Here we only sketch the way how it can be done.
First, one has to consider a particular coupling map between levels in given qudits.
A decomposition of single-qudit gates down to operations on allowed transitions have to be applied~\cite{Ringbauer2021, Mato2022}.
Second, a transformation of an employed two-qudit controlled-phase gate on particular native two-qudit operations is required. 
It can be realized via single-qudit gates.

Although manipulating with additional levels of qudits faces additional experimental problems, recent work (see, e.g., Refs.~\cite{OBrien2022, Ringbauer2021, Goss2022, Hill2021,Semerikov2022}) 
show a dramatic progress in increasing quality of operations with qudits based on various physical platforms.
Both single-qudit and two-qudit gates nowadays have achieved fidelities, which are comparable with fidelities, demonstrated on two-level systems.
We believe that combining these experimental achievements with the presented approach for decreasing the number of two-particle gates can significantly improve the resulting quality of quantum algorithms implementation.

\BLUE{
We also note that in order to extend our approach for three qubits embedded with a single ancillary level in a single qudit, one requires qudits of dimension $d=2^3+1=9$.
This is above typical dimensionality of currently considered qudit-based platforms~\cite{Ringbauer2021, Goss2022, Hill2021, Semerikov2022}.
Investigation of qudits with $d \geq 9$ together with possible accompanying practical challenges is an important promising topic for further research. 
}

\section*{Acknowledgments}

The research is supported by the Russian Science Foundation Grant No. 19-71-10091 (development of the method in Sec.~\ref{sec:Ququint} and Sec.~\ref{sec:Toffoli}) 
and the Priority 2030 program at the National University of Science and Technology ``MISIS'' under the project K1-2022-027 (application to Grover's algorithm in Sec.~\ref{sec:Grover}).

\bibliography{bibliography-qudits.bib}

\end{document}